\documentclass[showpacs,superscriptaddress,amsfonts,twocolumn]{revtex4}
\usepackage{bm}
\usepackage{times}
\usepackage[dvips]{graphicx}

\begin{document}

\title{Noise and Inertia-Induced Inhomogeneity in the Distribution of 
Small Particles in Fluid Flows}

\author{Julyan H. E. Cartwright} 
\homepage{http://lec.ugr.es/~julyan}
\email{julyan@lec.ugr.es}
\affiliation{Laboratorio de Estudios Cristalogr\'aficos, CSIC, 
E-18071 Granada, Spain}
\author{Marcelo O. Magnasco} 
\homepage{http://asterion.rockefeller.edu/marcelo/marcelo.html}
\email{marcelo@sur.rockefeller.edu}
\affiliation{Mathematical Physics Laboratory, The Rockefeller University, 
Box 212, 1230 York Avenue, NY 10021}
\author{Oreste Piro}
\homepage{http://www.imedea.uib.es/~piro}
\email{piro@imedea.uib.es} 
\affiliation{Institut Mediterrani d'Estudis Avan\c{c}ats, CSIC--UIB,
E-07071 Palma de Mallorca, Spain}

\date{version of \today}
 
\begin{abstract}
The dynamics of small spherical neutrally buoyant particulate impurities
immersed in a two-dimensional fluid flow are known to lead to particle
accumulation in the regions of the flow in which rotation dominates over shear,
provided that the Stokes number of the particles is sufficiently small. If the
flow is viewed as a Hamiltonian dynamical system, it can be seen that the
accumulations occur in the nonchaotic parts of the phase space: the
Kolmogorov--Arnold--Moser tori. This has suggested a generalization of these
dynamics to Hamiltonian maps, dubbed a bailout embedding. In this paper we use
a bailout embedding of the standard map to mimic the dynamics of impurities
subject not only to drag but also to fluctuating forces modelled as white
noise. We find that the generation of inhomogeneities associated with the
separation of particle from fluid trajectories is enhanced by the presence of
noise, so that they appear in much broader ranges of the Stokes number than
those allowing spontaneous separation.
\end{abstract}


\maketitle

\section{Introduction}

Impurities suspended in a fluid flow are frequently observed to be distributed
inhomogeneously. Even in very chaotic flows, particulate impurities arrange
themselves in extraordinarily structured distributions, in apparent
contradiction to the high mixing efficiency expected from the characteristics
of the basic flow. To give just one example, in the particular instance of
geophysical fluids, the filamentary structure, or patchiness, often displayed
by plankton populations in the oceans is a puzzling problem currently under
intense investigation \cite{levin,steele,abraham2}. Several mechanisms to
produce this type of inhomogeneity have been studied, and include both
dynamical aspects of the flow as well as the reactive properties of the
considered impurities. The basic idea in these mechanisms is that the particle
loss due either to the flow --- in open flows --- or to the chemical or
population dynamics of the particles --- in closed flows --- is minimized on
some manifolds associated with the hyperbolic character of the flow
\cite{jung,toroczkai,neufeld}. In this paper we explore an alternative purely
dynamical mechanism for inhomogeneity with non-reactive particles in bounded
flows. We show that particle inertial effects combined with fluctuating forces
are capable of producing inhomogeneity even in cases in which the impurity and
fluid densities match exactly.

When impurities have a different density to the fluid, it is intuitively clear
that they will be expelled from rapidly rotating regions of the flow --- for
heavy particles --- or attracted to the center of those regions --- for light
particles --- because of centrifugal effects \cite{partproc}.  However, it was
recently demonstrated that neutrally buoyant particles also tend to settle in
the rotation-dominated regions of a flow, but because of a more subtle
mechanism involving the separation between the fluid and particle trajectories
that can occur in the opposite regions, i.e., in the areas of the flow
dominated by shear \cite{neutpartprl}. However, this mechanism is only relevant
when the particle Stokes number is smaller than the eigenvalues of the Jacobian
matrix of the flow, a condition that may not be fulfilled in some physically
interesting situations. We show here by means of a minimal model that the
latter condition may be relaxed if a small amount of noise be added to the
forces acting on the impurity.

Our approach is qualitative, in the sense that instead of considering a
specific flow and the precise particle dynamics induced by it, we describe the
system with an iterative map whose evolution contains the basic features of
both the fluid flow and the particle dynamics: flow volume preservation,
together with particle separation at the hyperbolic regions. The reason for
moving to a discrete system is that in the map the phenomena that we describe
may be understood more intuitively, while translating the results back to the
flow case is immediate. The strategy of understanding fluid-dynamical phenomena
by using iterated maps amenable to the powerful artillery of dynamical-systems
theory has been successfully applied on several different occasions. For
example, the structures of the chaotic advection induced by time-periodic
three-dimensional incompressible flows were predicted by studying the
qualitatively equivalent dynamics of three-dimensional volume-preserving maps
\cite{feingold88.2,pirofein}. Later, these structures were confirmed in
realistic flows \cite{3dpaper,spheresletter,spherespaper}. Other examples are
the treatment of the propagation of combustion fronts in laminar flows by a
qualitative map approach \cite{abel}, and the description of the formation of
plankton population structures due to inhomogeneities of the nutrient sources
\cite{lopez}. Remarkably, in the instance we discuss in this paper, the
procedure is also useful from the point of view of dynamical-systems theory, as
it has suggested a new technique --- bailout embedding --- for the control of
Hamiltonian chaos \cite{bailout1}.

The plan of the paper is as follows. First, we briefly review the classical
model for the forces acting on a small spherical tracer moving relative to the
fluid in which it is immersed (Section~\ref{maxey}). Concentrating on the case
of a neutrally buoyant impurity, we trace the construction of a minimal model
that makes evident the separation of particle and fluid trajectories in the
regions in which the flow presents strong shear (Section~\ref{minimal}). On the
basis of this model, we make a generalization that allows us to build a
discrete mapping that represents the Lagrangian evolution of the fluid parcels
as well as the dynamics of the particle (Section~\ref{maps}). While this map
also displays particle--fluid separation when the parameter $\gamma$,
equivalent to the Stokes number, is relatively small, we show in
Section~\ref{noise} that a small amount of noise, added to the dynamics of the
particle to separate it continually from the flow, enhances the impact of the
hyperbolic regions far beyond the values of $\gamma$ required for separation.
Conclusions are to be found in Section~\ref{concs}.

\section{Maxey--Riley Equations}\label{maxey}

The equation of motion for a small, spherical tracer in an
incompressible fluid we term the Maxey--Riley equation
\cite{maxey,michaelides,neutpartprl}, which may be written as
\begin{eqnarray}\label{eom}
&\rho_p\displaystyle\frac{d\bm{v}}{dt}=&
\rho_f\displaystyle\frac{D\bm{u}}{Dt}+(\rho_p-\rho_f)\bm{g} \\
&&-\displaystyle\frac{9\nu\rho_f}{2a^2}
\left(\bm{v}-\bm{u}
-\displaystyle\frac{a^2}{6}\nabla^2\bm{u}\right) \nonumber \\
&&-\displaystyle\frac{\rho_f}{2}\left(\displaystyle\frac{d\bm{v}}{dt}-
\displaystyle\frac{D}{Dt}\left[\bm{u}
+\displaystyle\frac{a^2}{10}\nabla^2\bm{u}\right]\right) 
\nonumber \\
&&-\displaystyle\frac{9\rho_f}{2a}\displaystyle\sqrt\frac{\nu}{\pi}
\displaystyle\int_{0}^{t}
\displaystyle\frac{1}{\sqrt{t-\zeta}}\displaystyle\frac{d}{d\zeta}
\left(\bm{v}-\bm{u}
-\displaystyle\frac{a^2}{6}\nabla^2\bm{u}\right)d\zeta. \nonumber 
\end{eqnarray}
Here $\bm{v}$ represents the velocity of the particle, $\bm{u}$ that
of the fluid, $\rho_p$ the density of the particle, $\rho_f$, the density of 
the fluid it displaces, $\nu$, the kinematic viscosity of the fluid, $a$, the
radius of the particle, and $\mathbf g$, gravity. 
The derivative $D\bm{u}/Dt$
is along the path of a fluid element
\begin{equation}
\frac{D\bm{u}}{Dt}=\frac{\partial\bm{u}}{\partial t}
+(\bm{u}\cdot\bm{\nabla})\bm{u}
,\label{euler}\end{equation}
whereas the derivative $d\bm{u}/dt$ 
is taken along the trajectory of the particle
\begin{equation}
\frac{d\bm{u}}{dt}=\frac{\partial\bm{u}}{\partial t}
+(\bm{v}\cdot\bm{\nabla})\bm{u}
.\label{lagrange}\end{equation}
The terms on the right of Eq.\ (\ref{eom}) represent respectively the force
exerted by the undisturbed flow on the particle, buoyancy, Stokes drag, the
added mass due to the boundary layer of fluid moving with the particle
\cite{taylor,auton}, and the Basset--Boussinesq force
\cite{boussinesq,basset} that depends on the history of the relative
accelerations of particle and fluid. The terms in $a^2\nabla^2\bm{u}$ are
the Fax\'en \cite{faxen} corrections. The Maxey--Riley equation is derived
under the assumptions that the particle radius and its Reynolds number are
small, as are the velocity gradients around the particle.

\section{Minimal Model}\label{minimal}

First let us consider a minimal model for a neutrally buoyant particle. For this
we set $\rho_p=\rho_f$ in Eq.\ (\ref{eom}). We consider the Fax\'en
corrections and the Basset--Boussinesq term to be negligible. We now rescale
space, time, and velocity by scale factors $L$, $T=L/U$, and $U$, to arrive at
the expression
\begin{equation}
\frac{d\bm{v}}{dt}=\frac{D\bm{u}}{Dt}
-{\mathrm St}^{-1}\left(\bm{v}-\bm{u}\right) 
-\frac{1}{2}\left(\frac{d\bm{v}}{dt}-\frac{D\bm{u}}{Dt}\right)
,\label{neutral}\end{equation}
where ${\mathrm St}$ is the particle Stokes number
${\mathrm St}=2a^2 U/(9\nu L)=2/9\,(a/L)^2 {\mathrm Re}_f$, 
${\mathrm Re}_f$ being
the fluid Reynolds number. The assumptions involved in deriving 
Eq.\ (\ref{eom}) require that ${\mathrm St}\ll 1$ in Eq.\ (\ref{neutral}). 

If we substitute the expressions for the derivatives in
Eqs.\ (\ref{euler}) and (\ref{lagrange}) into Eq.\ (\ref{neutral}), we obtain
\begin{equation}
\frac{d}{dt}\left(\bm{v}-\bm{u}\right)=
-\left(\left(\bm{v}-\bm{u}\right)\cdot\bm{\nabla}\right)\bm{u}
-\frac{2}{3}\,{\mathrm St}^{-1}\left(\bm{v}-\bm{u}\right) 
.\end{equation}
We may then write $\bm{A}=\bm{v}-\bm{u}$, whence 
\begin{equation}
\frac{d\bm{A}}{dt}=-\left(J+\frac{2}{3}\,{\mathrm St}^{-1} {\mathrm I}\right)
\cdot\bm{A}
,\label{Aeqn}\end{equation}
where $J$ is the Jacobian matrix  --- we now concentrate on two-dimensional
flows $\bm{u}=(u_x, u_y)$ ---
\begin{equation}
J=
\left(
\begin{array}{cc}
\partial_x u_x & \partial_y u_x \\
\partial_x u_y & \partial_y u_y
\end{array}
\right)
.\end{equation}
If we diagonalize the matrix we obtain
\begin{equation}
\frac{d\bm{A}_D}{dt}=
\left(
\begin{array}{cc}
\lambda-2/3\,{\mathrm St}^{-1} & 0 \\
0 & -\lambda-2/3\,{\mathrm St}^{-1}
\end{array}
\right)
\cdot\bm{A}_D
,\label{AD}\end{equation}
so if ${\mbox{\it Re}}(\lambda)>2/3\,{\mathrm St}^{-1}$, $\bm{A}_D$ may 
grow exponentially. Now $\lambda$ satisfies 
$\mathrm{det}(J-\lambda {\mathrm I})=0$, so 
$\lambda^2-{\mathrm tr}J+{\mathrm det}J=0$. Since the flow is incompressible,
$\partial_x u_x+\partial_y u_y=\mathrm{tr}J=0$,
thence $-\lambda^2=\mathrm{det}J$. Given squared vorticity  
$\omega^2=(\partial_x u_y-\partial_y u_x)^2$, 
and squared strain $s^2=s_1^2+s_2^2$, where the normal component is
$s_1=\partial_x u_x-\partial_y u_y$ and the shear component is
$s_2=\partial_y u_x+\partial_x u_y$, we may write
\begin{equation}
Q=\lambda^2=-{\mathrm det}J=(s^2-\omega^2)/4
,\end{equation}
where $Q$ is the Okubo--Weiss parameter \cite{okubo,weiss2}.
If $Q>0$, $\lambda^2>0$, and $\lambda$ is real, so deformation dominates, as
around hyperbolic points, whereas if $Q<0$, $\lambda^2<0$, and $\lambda$ is
complex, so rotation dominates, as near elliptic points. Equation (\ref{Aeqn})
together with $d\bm{x}/dt=\bm{A}+\bm{u}$ defines a dissipative
dynamical system 
\begin{equation} 
d\mbox{\boldmath$\xi$}/dt=\bm{F}(\mbox{\boldmath$\xi$}) 
\label{disssys}
\end{equation} 
with constant divergence $\bm{\nabla}\cdot\bm{F}=-4/3\,{\mathrm St}^{-1}$ in 
the four dimensional phase space $\mbox{\boldmath$\xi$}=(x,y,A_x,A_y)$, so that
while small values of ${\mathrm St}$ allow for large values of the divergence, 
large values of ${\mathrm St}$ force the divergence to be small. The Stokes
number is the relaxation time of the particle back onto the fluid trajectories 
compared to the time scale of the flow --- with larger ${\mathrm St}$, the 
particle has more independence from the fluid flow. From Eq.\ (\ref{AD}), about 
areas of the flow near to hyperbolic stagnation points with 
$Q>4/9\,{\mathrm St}^{-2}$, particle and flow trajectories separate 
exponentially.

\begin{figure}[tbhp]
\begin{center}
\includegraphics*[width=0.49\columnwidth]{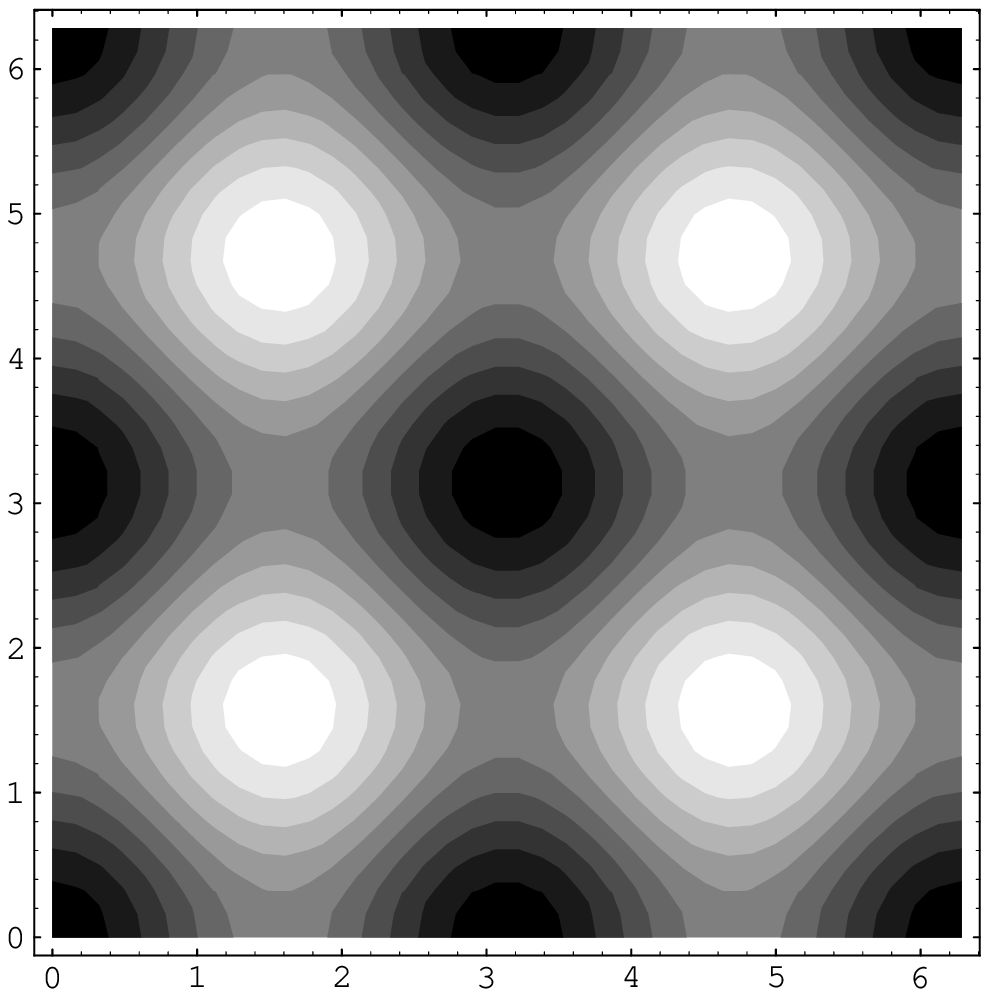}
\includegraphics*[width=0.49\columnwidth]{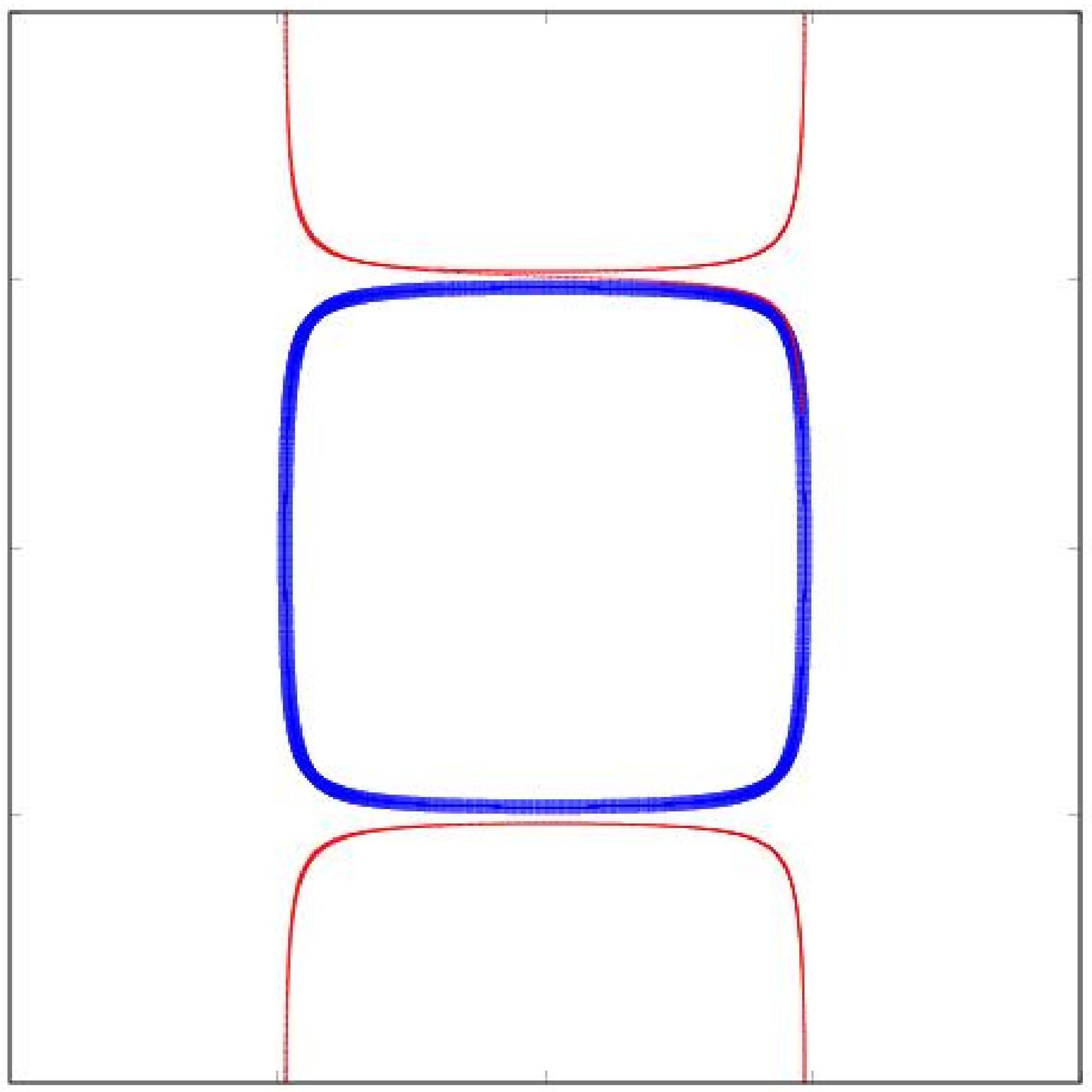}\\
\includegraphics*[width=0.49\columnwidth]{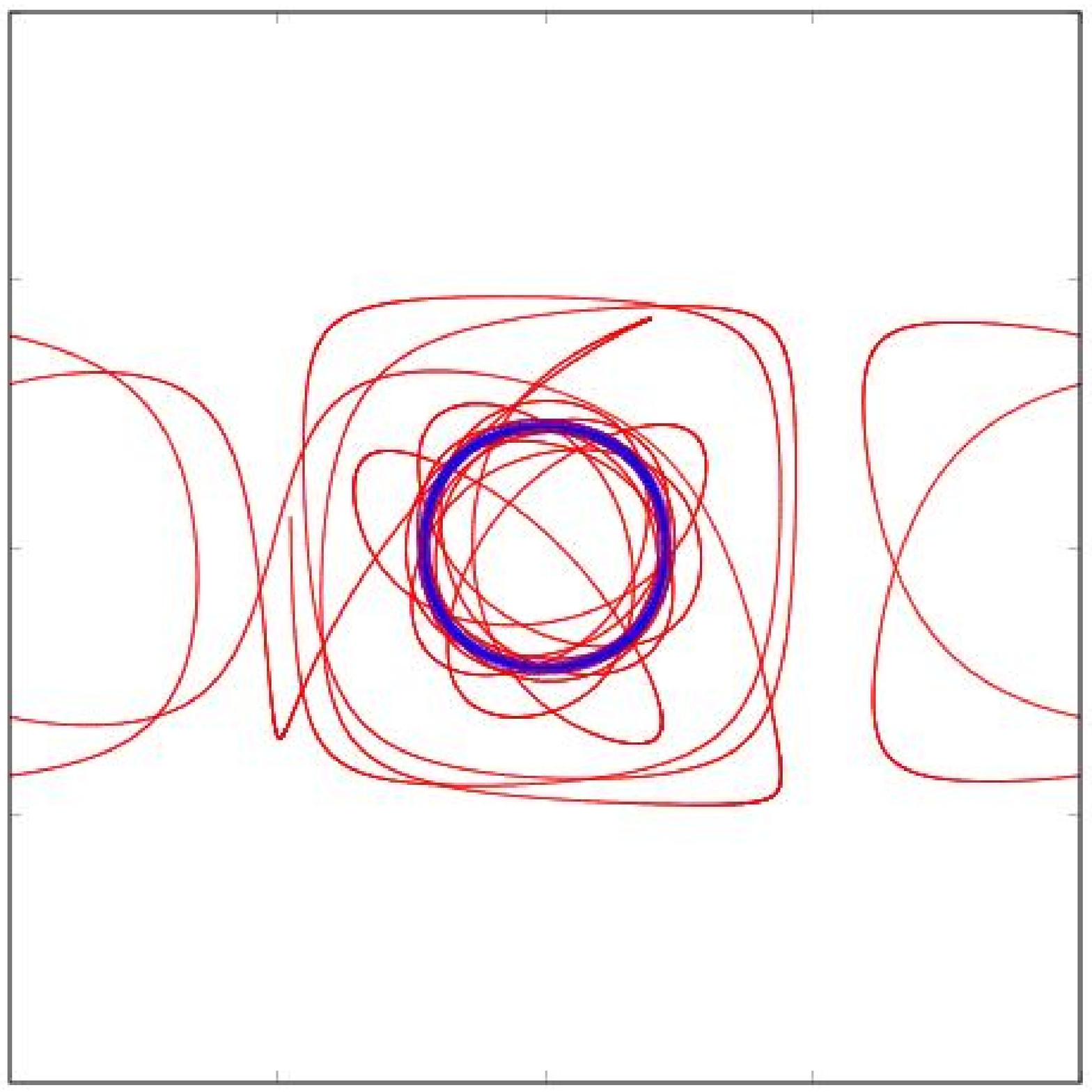}
\includegraphics*[width=0.49\columnwidth]{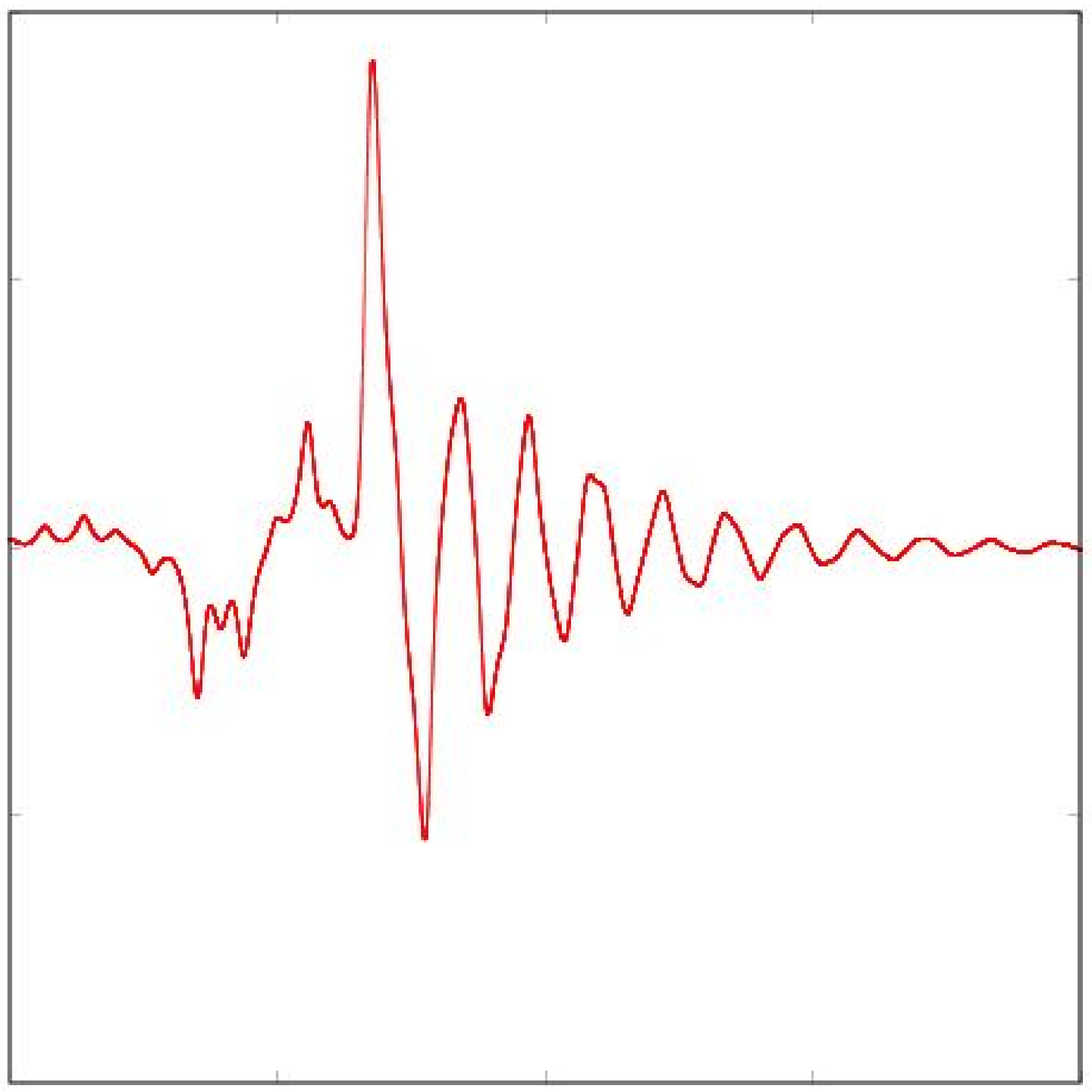}
\end{center}
\caption{\label{timeind}
(a) Contour plot illustrating magnitude of $Q$ --- lighter is higher $Q$ 
--- for the time-independent model Eq.\ (\ref{flow}) (the flow is on a torus).
(b) The separation of a neutrally buoyant particle trajectory 
(red line) from the flow (blue line) in regions of high $Q$ allows the 
particle to wander between cells.
(c) After a complicated excursion, a particle (red line) eventually 
settles in a zone of low $Q$ of the flow (blue line).
(d) The velocity difference $v_x - u_x$ between the particle and the 
flow against time.
}
\end{figure}

To illustrate the effects of ${\mathrm St}$ and $Q$ on the dynamics of a 
neutrally buoyant particle, let us consider the simple incompressible
two-dimensional model flow defined by the stream function
\begin{equation}
\psi(x,y,t)=A \cos(x+B\sin\omega t)\cos y
.\label{flow}\end{equation}
The equations of motion for an element of the fluid will then be
$\dot{x}=\partial_y\psi$, $\dot{y}=-\partial_x\psi$. 
$\psi$ has the r\^ole of a Hamiltonian for the dynamics of such an element,
with $x$ and $y$ playing the parts of the conjugate coordinate and momentum
pair. Let us first consider the simplest case, for which we suppress time 
dependence by setting $B=0$. Thence $\psi$ should be a constant of motion,
which implies that real fluid elements follow trajectories that are level
curves of $\psi$. In Fig.~\ref{timeind}(a) are depicted contours of $Q$.  Notice
that the high values of $Q$ are around the hyperbolic points, while negative
$Q$ coincides with the centres of vortices --- elliptic points --- in the flow.
Figure~\ref{timeind} (top right)  shows the trajectory of a neutrally buoyant
particle starting from a point on a fluid trajectory within the central vortex,
but with a small velocity mismatch with the flow. This mismatch is amplified in
the vicinity of the hyperbolic stagnation points for which $Q$ is larger than
$4/9\,{\mathrm St}^{-2}$ to the extent that the particle  leaves the central
vortex for one of its neighbors, a trip that is not allowed to a fluid parcel.
In the end a particle settles on a trajectory that does not visit regions of
high $Q$, proper for a fluid parcel. While this effect is already seen 
in Fig.~\ref{timeind}(b), it is more dramatically pictured in the
trajectory shown in Fig.~\ref{timeind}(c), in which the particle performs a long and
complicated excursion wandering between different vortices before it settles in
a region of low $Q$ of one of them. To illustrate the divergence of particle
and fluid trajectories, and the fact that particle and fluid finally arrive at
an accord, in Fig.~\ref{timeind}(d) we display the difference between the particle
velocity and the fluid velocity at the site of the particle against time for
this case. Notice that this difference seems negligible at time zero, and that
it also convergences to zero at long times, but during the interval in which
the excursion takes place it fluctuates wildly. 

\begin{figure}[tbhp]
\begin{center}
\includegraphics*[width=0.49\columnwidth]{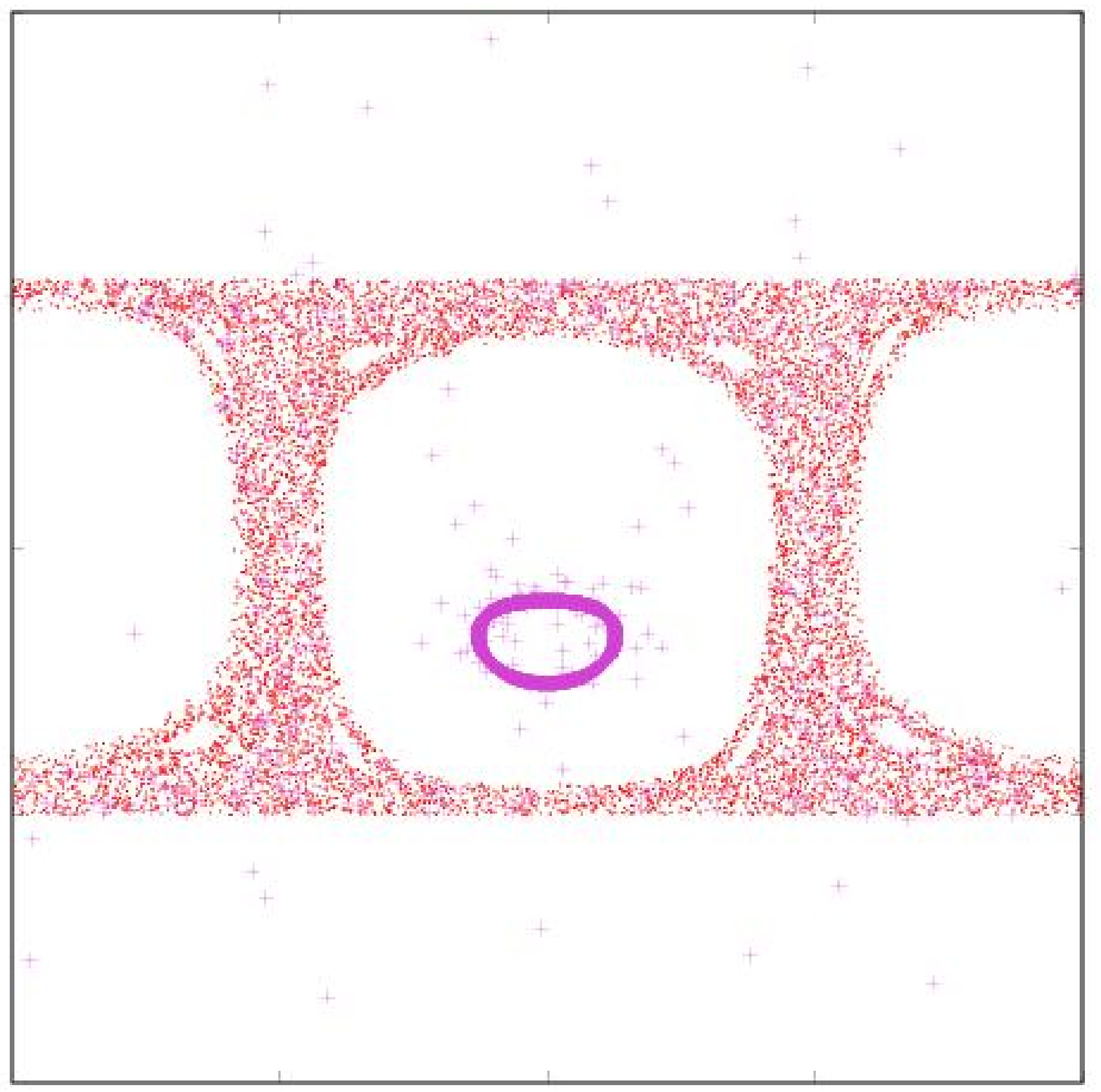}
\includegraphics*[width=0.49\columnwidth]{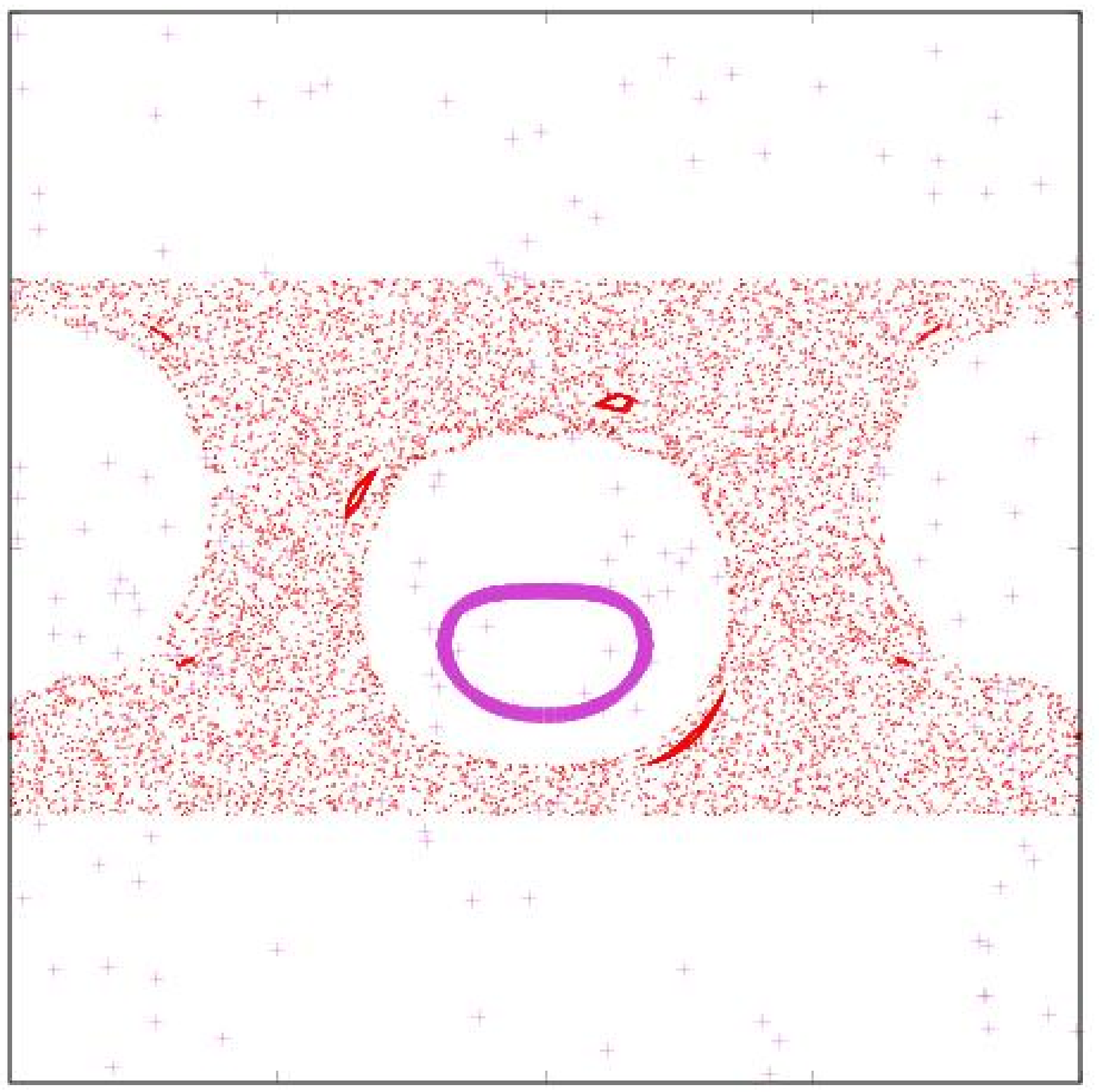} \\
\includegraphics*[width=0.49\columnwidth]{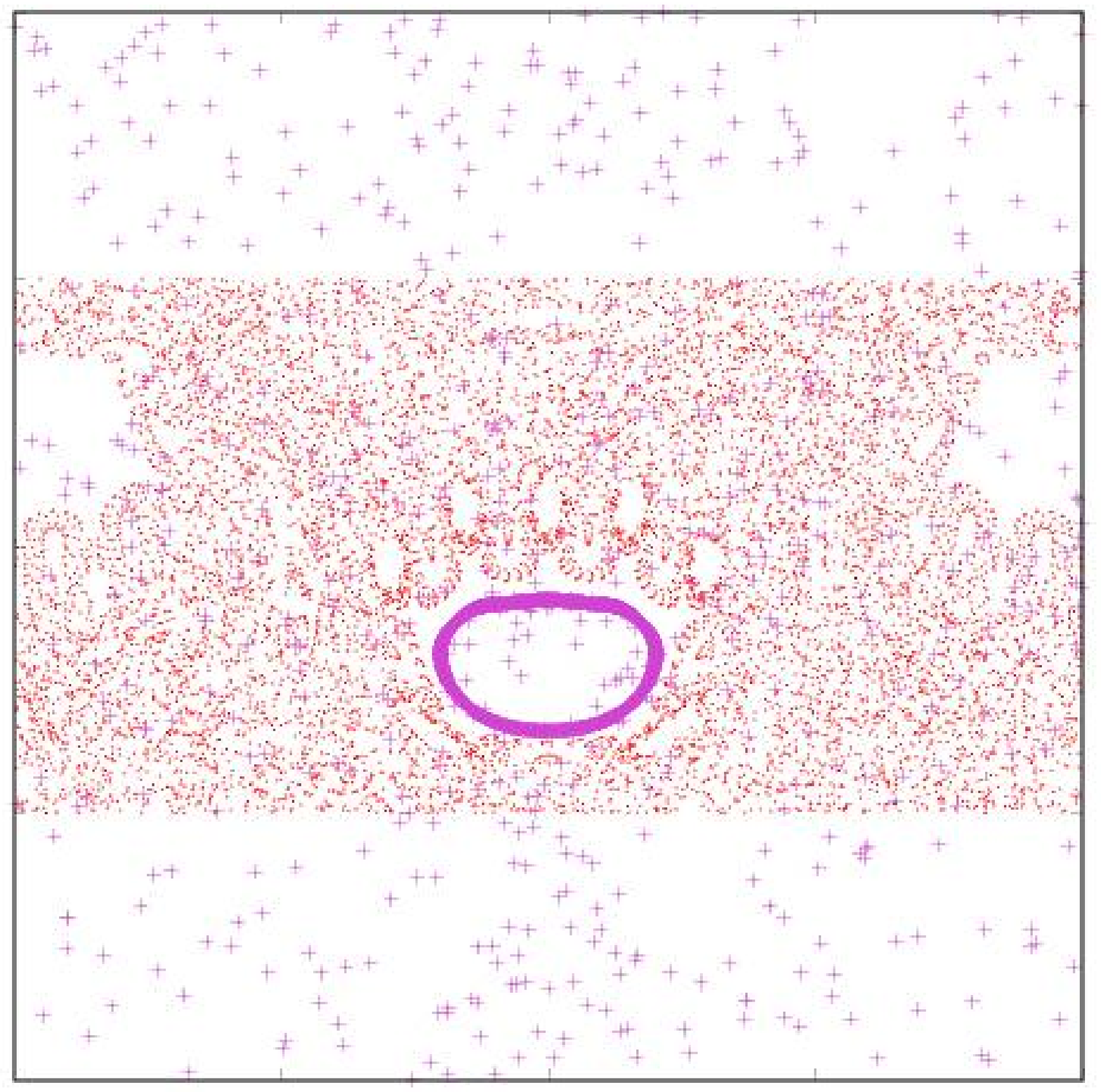}
\includegraphics*[width=0.49\columnwidth]{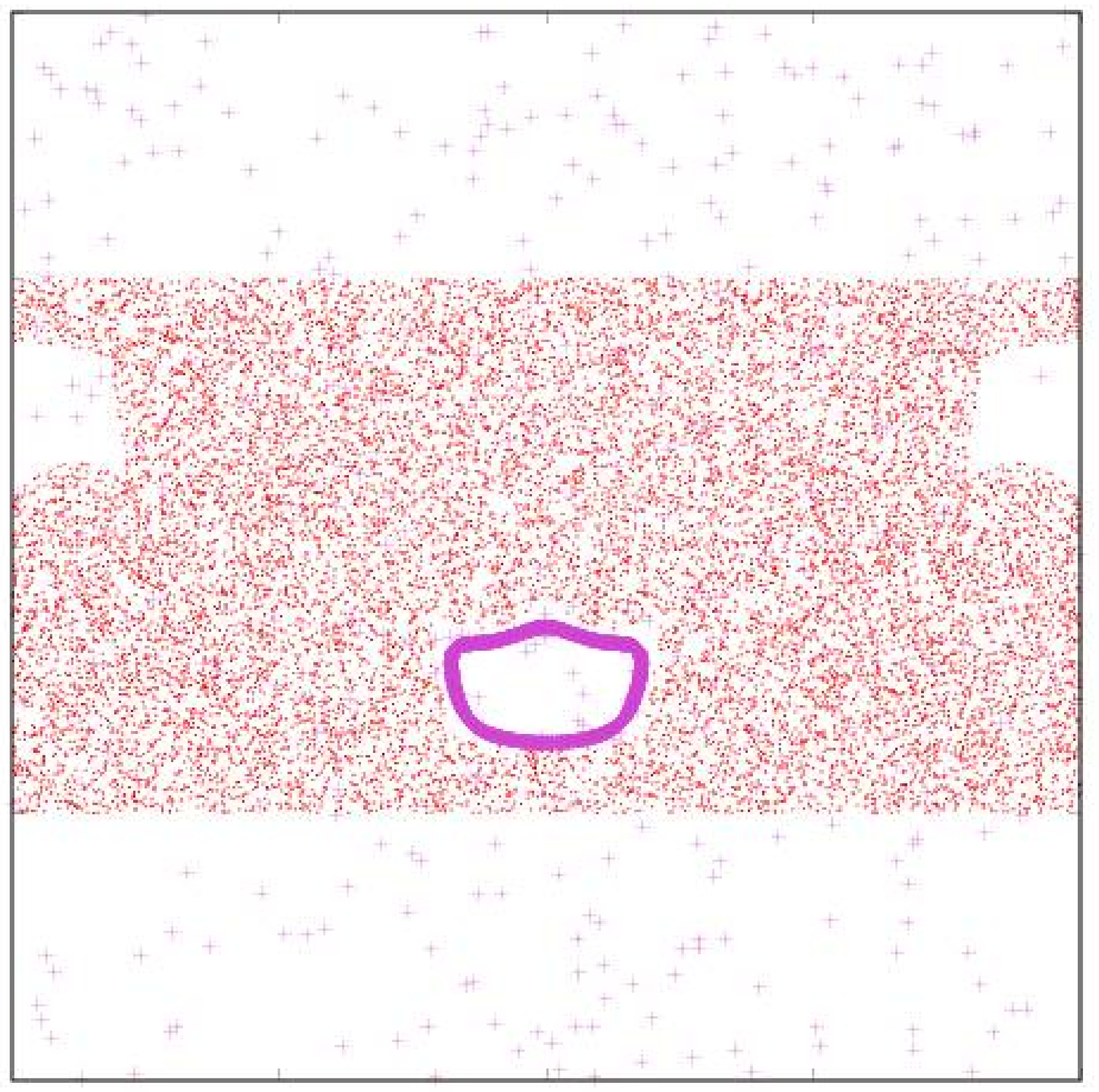}
\end{center}
\caption{\label{timedep}
Poincar\'e sections of trajectories in the time-dependent flow of Eq.\
(\ref{flow}). From top left to bottom right are shown (red dots) four
increasingly chaotic examples of the flow, and (magenta crosses), the
trajectories of neutrally buoyant particles in the flows that in each case
finally end up on a KAM torus within the regular region of the flow.}
\end{figure}

Even more interesting is the case of time-dependent flow: $B\neq 0$ in our
model. As in a typical Hamiltonian system, associated with the original
hyperbolic stagnation points, there are regions of the phase space ---  here
real space --- dominated by chaotic trajectories. Trajectories of this kind,
stroboscopically sampled at the frequency of the flow, are reproduced in
Fig.~\ref{timedep}. Such trajectories visit a large region of the space, which
includes the original hyperbolic stagnation points and their vicinities where
$Q$ is large. Excluded from the reach of such a chaotic trajectory remain areas
where the dynamics is regular; the so-called KAM (Kolmogorov--Arnold--Moser)
tori. In our model these lie in the regions where $Q<4/9\,{\mathrm St}^{-2}$. 
Now a neutrally buoyant particle trying to follow a chaotic flow pathline would
eventually reach the highly hyperbolic regions of the flow. This makes likely
its separation and departure from such a pathline, in search of another
pathline to which to converge. However, convergence will only be achieved if
the pathline never crosses areas of high $Q$. Figure~\ref{timedep} demonstrates
this phenomenon: particles were released in the chaotic zone with a small
velocity mismatch. Such a particle follows the flow, until, coming upon a
region of sufficiently high $Q$, it is thrown out of that flow pathline onto a
long excursion that finally ends up in a regular region of the flow on a KAM
torus. The regular regions of the flow then constitute attractors of the
dissipative dynamical system Eq.\ (\ref{disssys}) that describes the behaviour
of a neutrally buoyant particle.

\section{Discrete Dynamics Description}\label{maps}

An examination of the dynamical system defining our minimal model for the
behavior of neutrally buoyant particles shows that it is composed of some
dynamics within some other larger set of dynamics. Eq.~(\ref{neutral}) can be
seen as an equation for a variable that would define another equation of
motion, when the solution of the former be zero. Recall that the inner dynamics
is that representing the Lagrangian trajectories of the flow, and that in two
dimensions, it is a Hamiltonian system. We may say that this Hamiltonian system
is embedded in a larger dynamical system, this time dissipative, whose
trajectories may or may not converge to zero. If they do, the system ends up
on the same trajectories as those of the smaller embedded system, but in general
this need not be the case. The general concept of some dynamics embedded within
some other dynamics may be exploited within the framework of dynamical-systems
theory to design techniques that reject some unwanted trajectories of the
original dynamics by making them unstable in the embedding, paralleling what
the particles do in the fluid-dynamics case. This idea has been dubbed bailout
embedding of a dynamical system and has obvious applications to control theory
\cite{bailout1}.

Trading generality for clarity, in this Section we present an example of a
bailout embedding for discrete-time dynamical sytems that closely represents
all the qualitative aspects of the above described impurity dynamics. 
Generally speaking, given a map 
$\bm{x}_{n+1}=\bm{T}(\bm{x}_{n})$ --- $\bm{x}$ being a
point in a space of arbitrary dimension --- a bailout embedding is the second
order recurrence 
\begin{equation}
\bm{x}_{n+2}-\bm{T}(\bm{x}_{n+1})=
{\mathrm K}(\bm{x}_n)(\bm{x}_{n+1}-\bm{T}(\bm{x}_{n}))
,\label{mapembed}\end{equation}
where ${\mathrm K}(\bm{x})$ is chosen such that $|{\mathrm K}(\bm{x})|>1$ over the
unwanted set of orbits, so that they become unstable in the embedding. In the
discrete system, almost any expression written for the ordinary differential
equation translates to something close to an exponential; in particular,
stability eigenvalues have to be negative in the ordinary-differential-equation
case to represent stability, while they have to be smaller than one in absolute
value in the map case. In order to simulate the dynamics of particles, the
operator $-\left(J+2/3\,{\mathrm St}^{-1} {\mathrm I}\right)$ should translate into the 
particular choice 
\begin{equation}
{\mathrm K}(\bm{x})=e^{-\gamma}\bm{\nabla}\bm{T}
\label{kexpress}\end{equation}
with $\gamma = -{2/3}\,{\mathrm St}^{-1} $, in the map setting.

To represent qualitatively a chaotic two-dimensional incompressible base flow 
we choose a classical testbed of Hamiltonian systems, the area preserving
standard map introduced by Taylor and Chirikov:
\begin{eqnarray}
\bm{T}:(x_n,y_n) \to (x_{n+1},y_{n+1})
,\label{smabstract}\end{eqnarray}
where
\begin{eqnarray}
x_{n+1} & = & x_{n}+\frac{k}{2\pi}\sin (2\pi y_{n}) \nonumber \\
y_{n+1} & = & y_{n}+x_{n+1}
\label{smexplicit}\end{eqnarray}
and $k$ is the parameter controlling integrability. Recall that in general, the
dynamics defined by this map present a mixture of quasiperiodic motions
occurring on the KAM tori and chaotic ones, depending on where we choose the
initial conditions. As the value of $k$ is increased, the region dominated by
chaotic trajectories pervades more and more of the phase space, except for
increasingly small islands of KAM quasiperiodiciy. Hence our qualitative
description of the impurity dynamics, the bailout embedding of the standard
map, is given by the coupled second-order iterative system defined by
Eqs.~(\ref{mapembed}), (\ref{kexpress}), (\ref{smabstract}), and
(\ref{smexplicit}).

Notice that due to the area-preserving property of the standard map, the two
eigenvalues of the derivative matrix must multiply to one. If they are complex,
this means that both have an absolute value of one, while if they are real,
generically one of them will be larger than one and the other smaller. We can
then separate the phase space into elliptic and hyperbolic regions
corresponding to each of these two cases. If a trajectory of the original map
lies entirely on the elliptic regions, the overall factor $\exp(-\gamma)$ would
damp any small perturbation away from it in the embedded system. But for
chaotic trajectories that inevitably visit some hyperbolic regions, there 
exists a value of $\gamma$ such that perturbations away from a standard-map 
trajectory are amplified instead of dying out in the embedding. As a
consequence, trajectories are expelled from the chaotic regions finally to
settle in the elliptic KAM islands.

\begin{figure}[tbhp]
\begin{center}
\includegraphics[width=0.49\columnwidth]{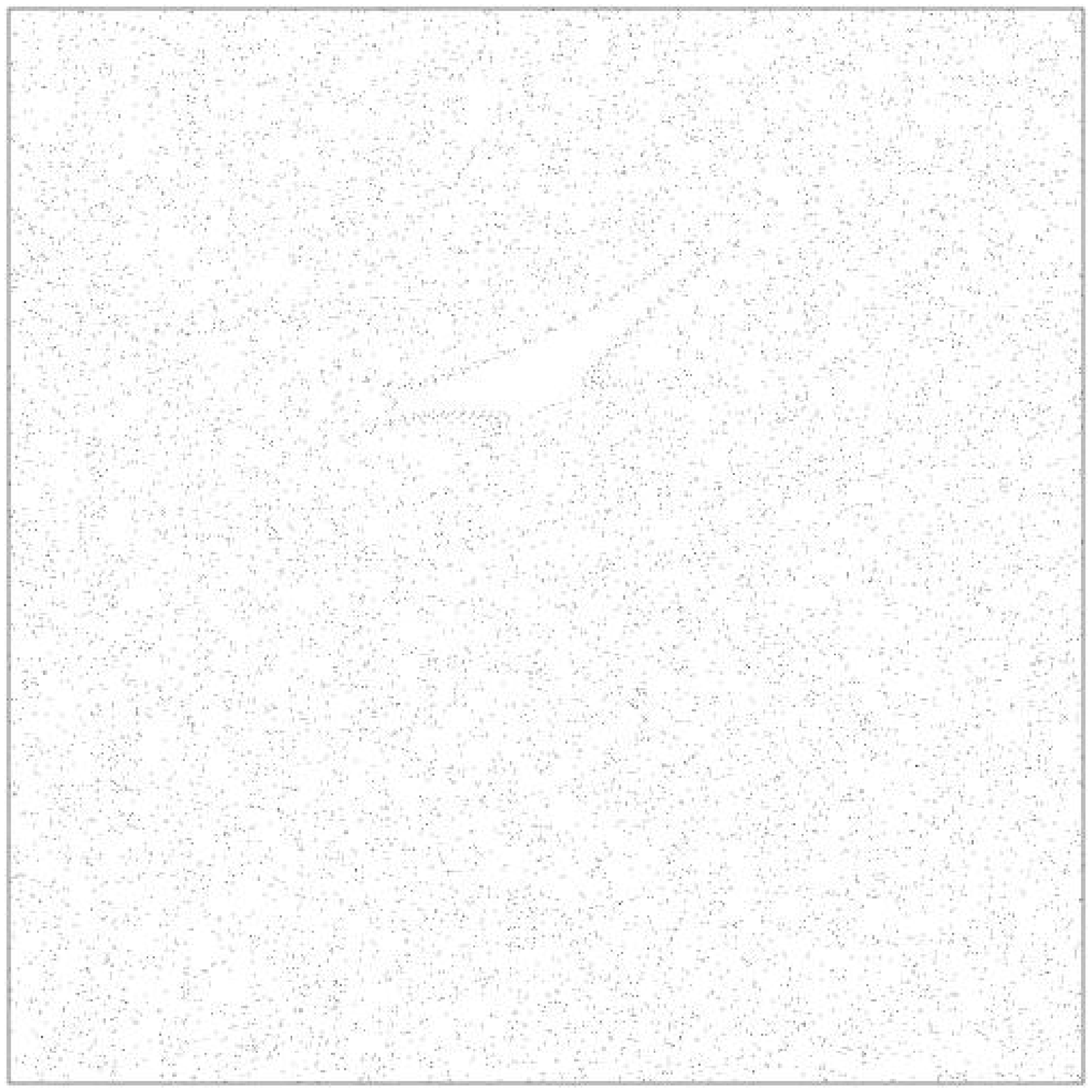}
\includegraphics[width=0.49\columnwidth]{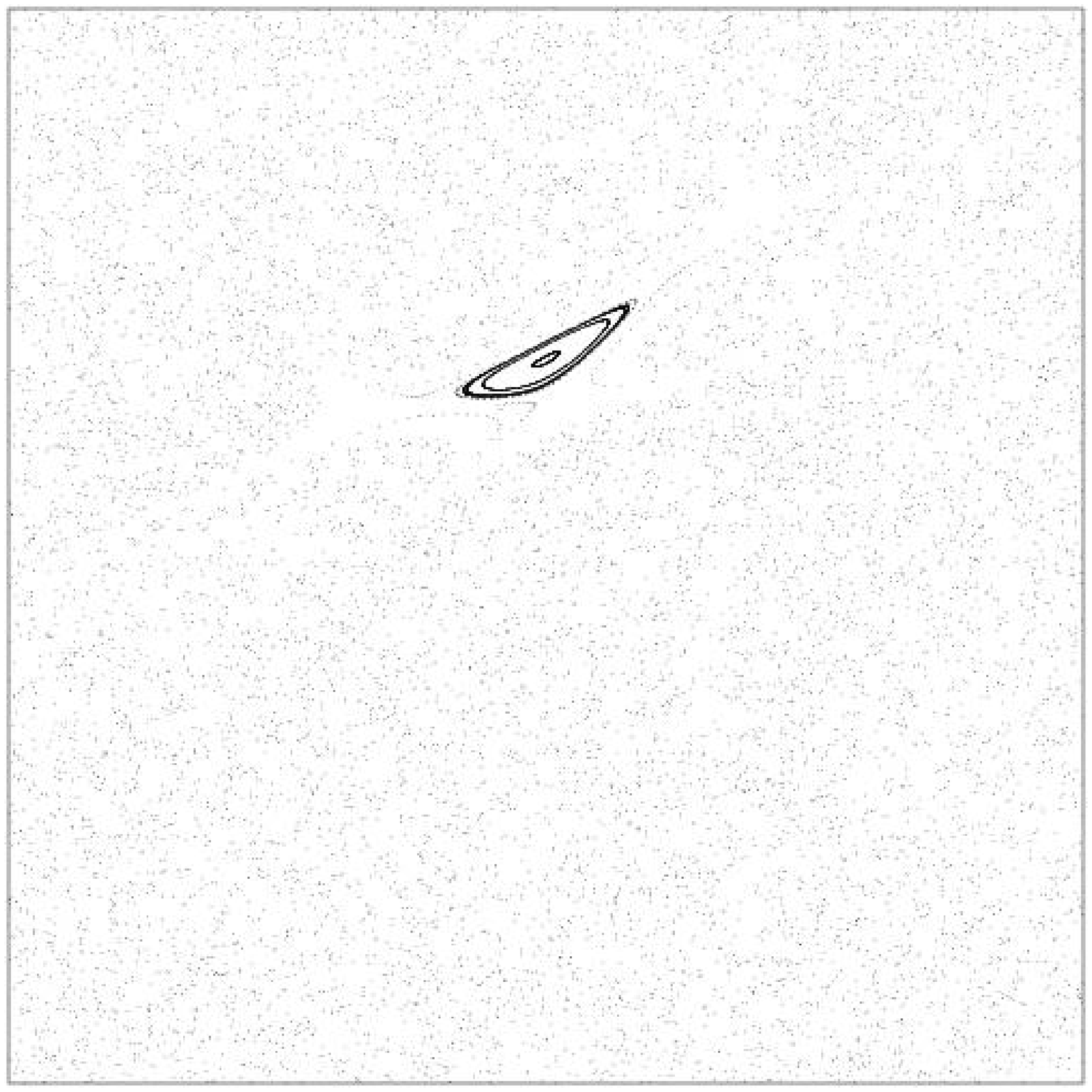} \\
\includegraphics[width=0.49\columnwidth]{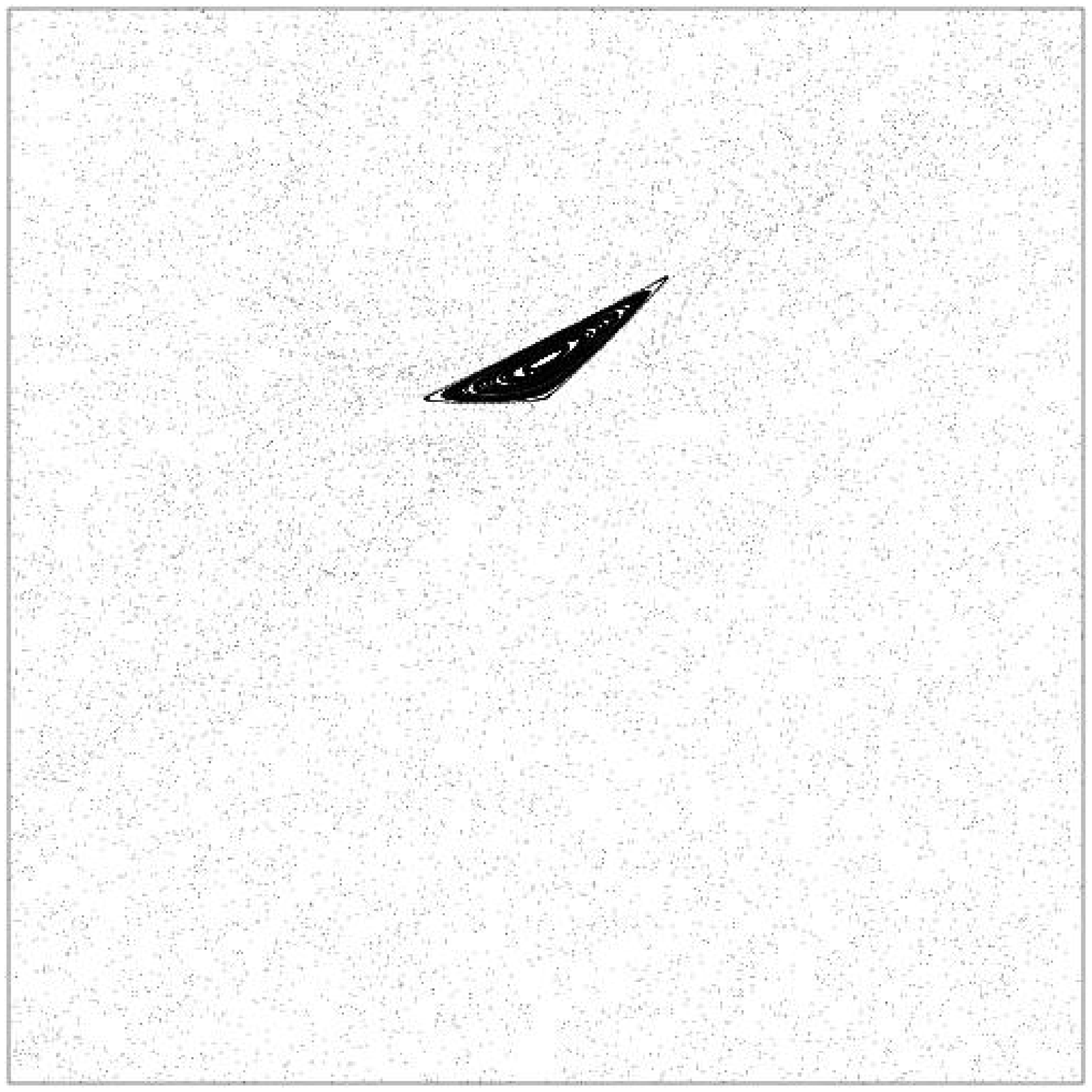}
\includegraphics[width=0.49\columnwidth]{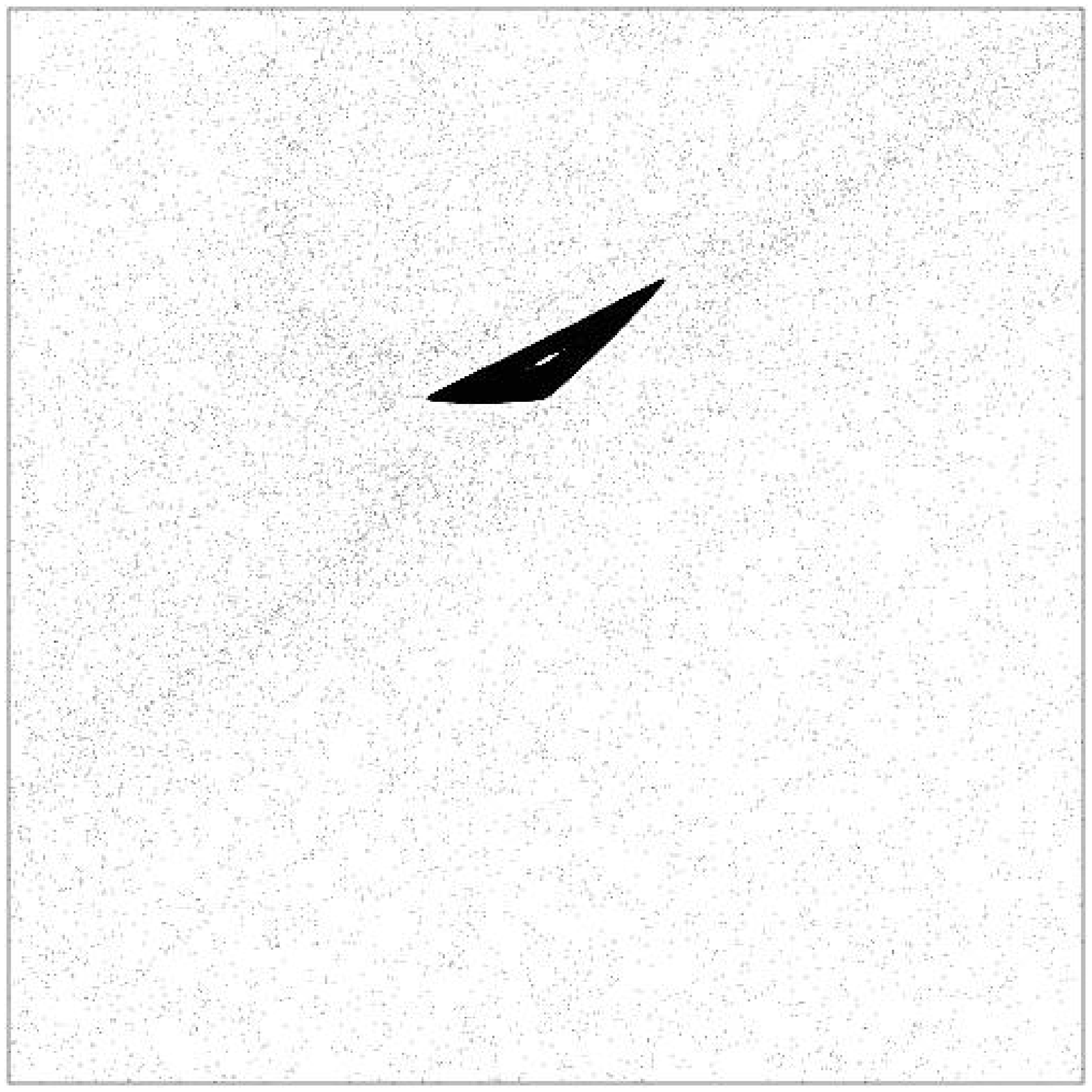}
\end{center}
\caption{\label{k=7}
The standard map for $k=7$ has a chaotic sea covering almost the entire torus,
except for a tiny period-two KAM island near position $0,2/3$.  1000 random
initial conditions were chosen, iterated for 20000 steps, then the next 1000
iterations are shown. The images here are a box $-0.05<x<0.05$,
$0.61<p<0.71$.
(a) Original map, (b) $\gamma=1.4$, (c) $\gamma=1.3$, (d) $\gamma=1.2$. 
}
\end{figure}

To illustrate this, Fig.~\ref{k=7} shows the phenomenon in a situation in which
the nonlinearity parameter has been set to the value $k=7$. This corresponds to
a very chaotic region of the standard map, characterized by the existence of
minute KAM islands within a sea of chaos that covers almost all the available
phase space. Fig.~\ref{k=7}(a) is a close-up --- to make it visible --- of the
largest of these islands and the dots there represent the successive positions
of a set of 1000 fluid parcels --- evolving according to the standard map ---
spread initially at random over the unit cell. Since none of these parcels were
initially located inside the island, this is seen as a white spot never visited
by the parcels. In contrast, in Figs~\ref{k=7}(b), \ref{k=7}(c), and
\ref{k=7}(d), the dots are the successive positions, after a number of
equilibration iterations, of particles initially placed as the parcels in
Fig.~\ref{k=7}(a), but allowing a very small initial discrepancy
$\bm{\delta}_0 = \bm{x}_1 - \bm{T}(\bm{x}_0)$ of both
dynamics. Notice that now, although having started initially outside the
island, some of the particles settle inside in a process that becomes
increasingly marked as the parameter $\gamma$ decreases.

\section{Noisy Dynamics}\label{noise}

By virtue of volume preservation, the invariant measure of the fluid-parcel
dynamics is either uniform, if the system is ergodic, or else disintegrates into a
foliation of KAM tori and ergodic regions, otherwise. In any case, the
addition of a small amount of white noise, which may be considered to represent
the effects of small scale turbulence, thermal fluctuations, etc., renders the
system ergodic with only a uniform invariant measure. Thus, the distribution of
fluid particles is expected to be uniform with or without the presence of
noise. The situation is however very different if the noise is applied to the
dynamics of the particles, or, correspondingly, to the bailout embedding.

Let us consider the following stochastic discrete-time dynamics
\begin{eqnarray}
\bm{x}_{n+2}-\bm{T}(\bm{x}_{n+1})&=&e^{-\gamma}\bm{\nabla}
   \bm{T}|_{\bm{x}_{n}}(\bm{x}_{n+1}-\bm{T}
   (\bm{x}_{n})) \nonumber \\
&&+\bm{\xi}_{n}
,\end{eqnarray}
in which, as in the previous Section, $\bm{x}$ represents the
particle coordinates and $\bm{T}(\bm{x})$ the fluid parcel evolution.
New here is the noise term $\xi_{n}$, with statistics
\begin{eqnarray}
\langle {\bm{\xi}_{n}} \rangle &=& 0, \nonumber \\
\langle {\bm{\xi}_{n}\,\bm{\xi}_{m}} \rangle &=&
\varepsilon(1-e^{-2\gamma})\,\delta_{m.n}\,{\mathrm I}
.\end{eqnarray}
This term forces the particle away from the fluid
trajectory at every step of the dynamics. However, the actual magnitude
of the fluctuations induced in $\bm{x}$ will be modulated by the
properties of $\bm{\nabla}\bm{T}$ --- the flow gradients --- along the 
particle trajectory.
Notice that for practical reasons we vary the noise intensity
in correspondence with $\gamma$ in order to obtain comparable
fluctuations at different values of this parameter.

Let us again consider the standard map as
modelling the basic flow, and its noisy bailout embedding that represents a
particulate impurity subject to both fluid drag and noise forces. We are
interested in the asymptotic stationary behavior of an ensemble of such
particles which, invoking the ergodicity of the fluctuations, should be well
represented by the the histogram of visits that a single particle pays to each
bin of the space --- the full phase space for the basic flow, but
a projection of the full phase space for the particles --- as time goes by.

\begin{figure}[tbhp]
\begin{center}
\includegraphics[width=0.49\columnwidth]{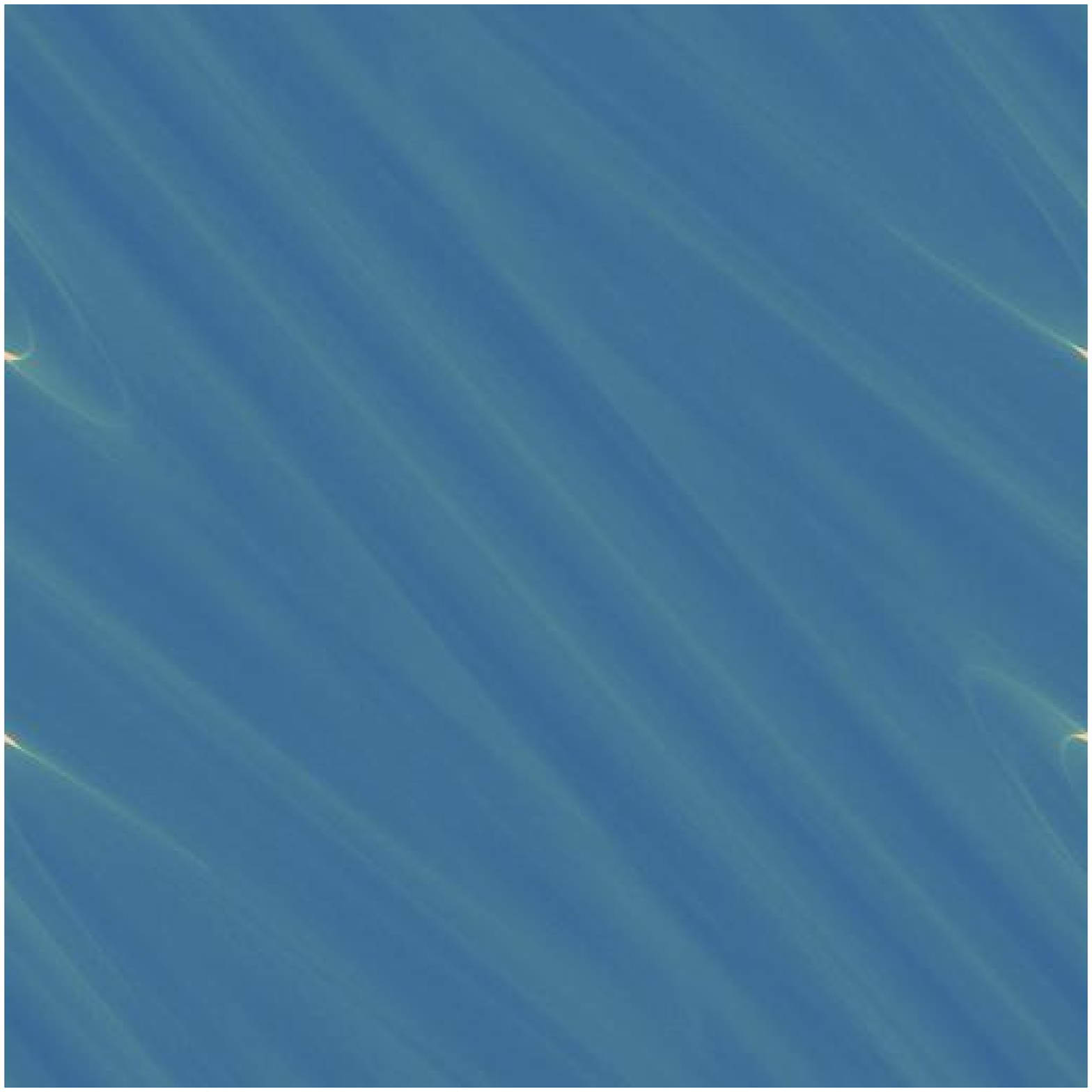}
\includegraphics[width=0.49\columnwidth]{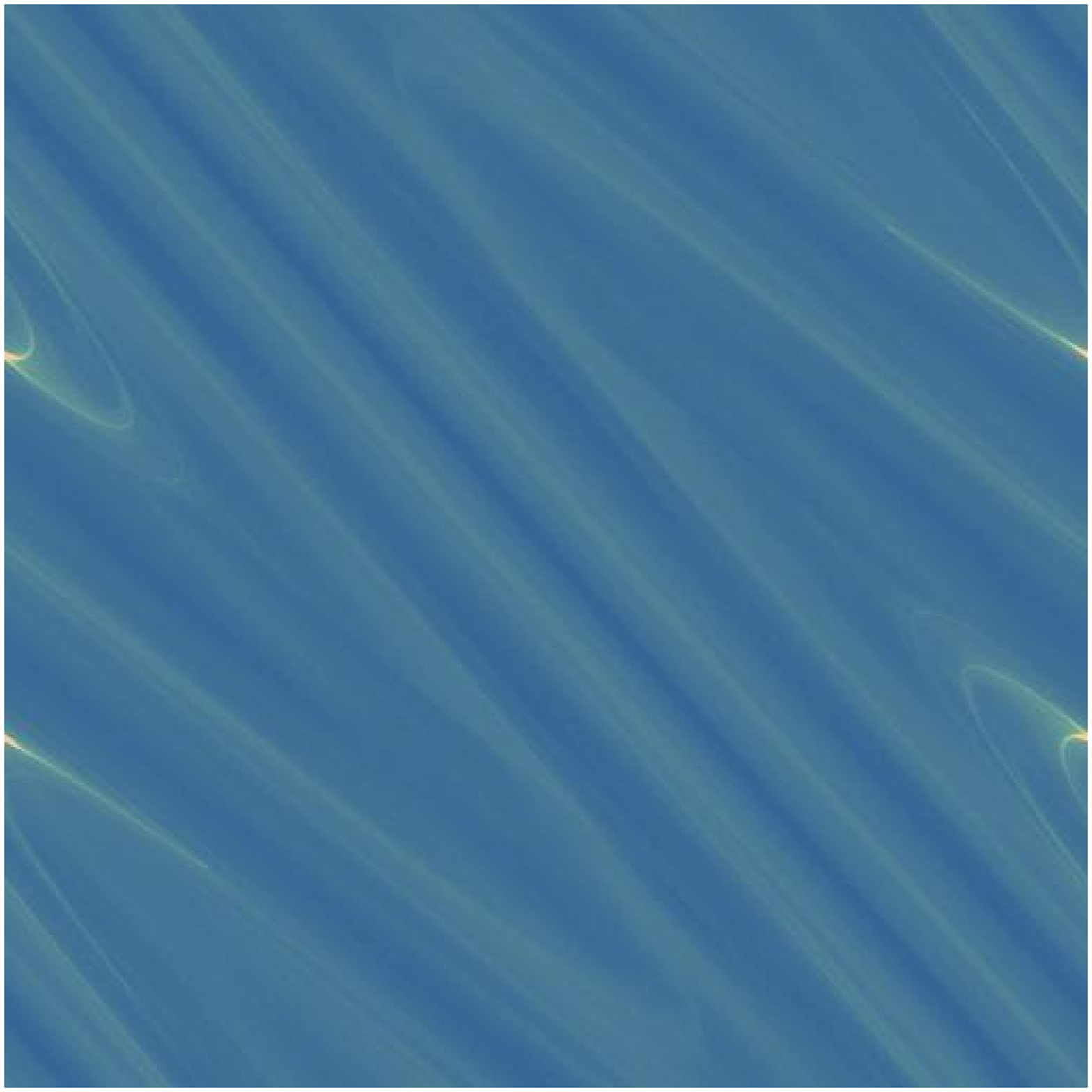} \\
\includegraphics[width=0.49\columnwidth]{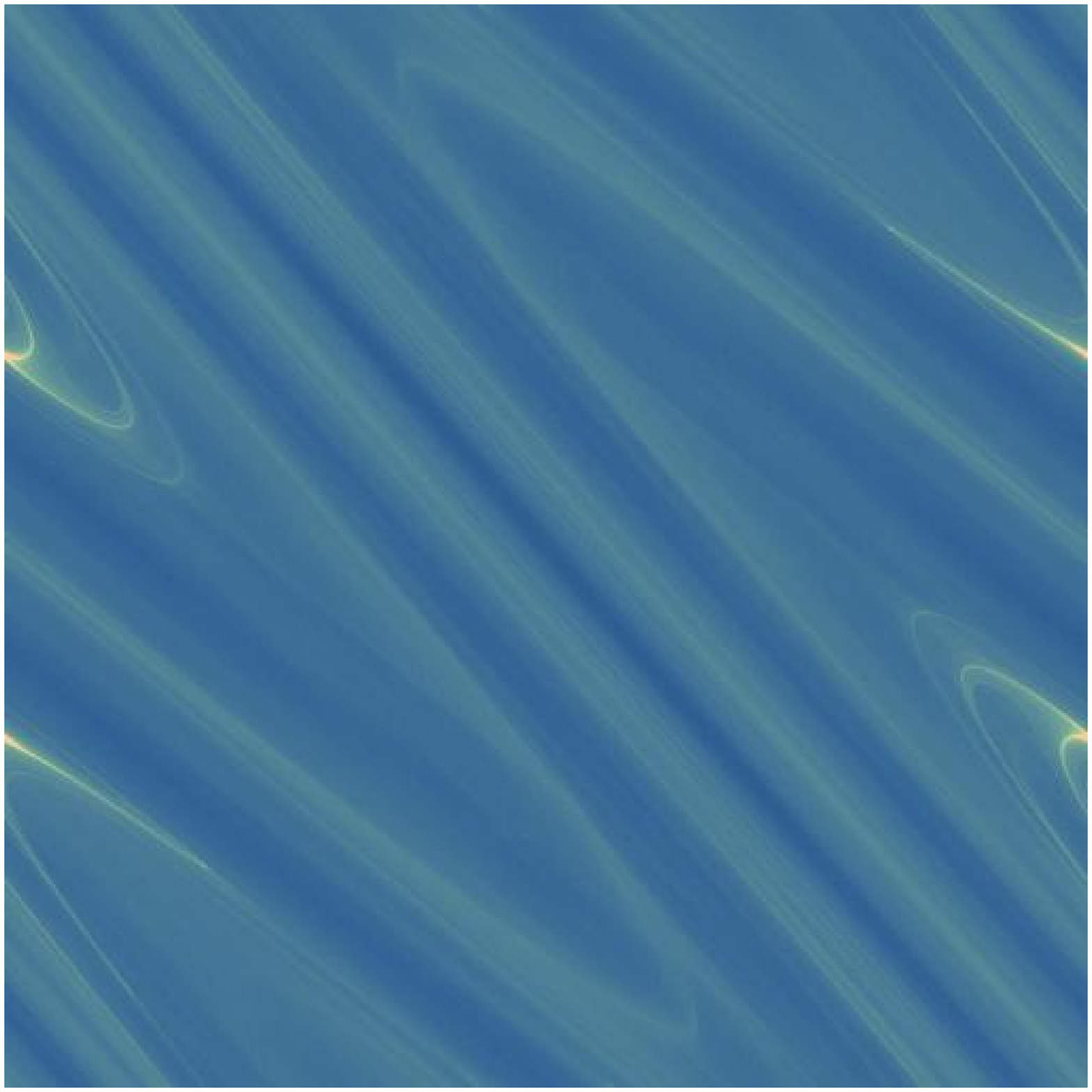}
\includegraphics[width=0.49\columnwidth]{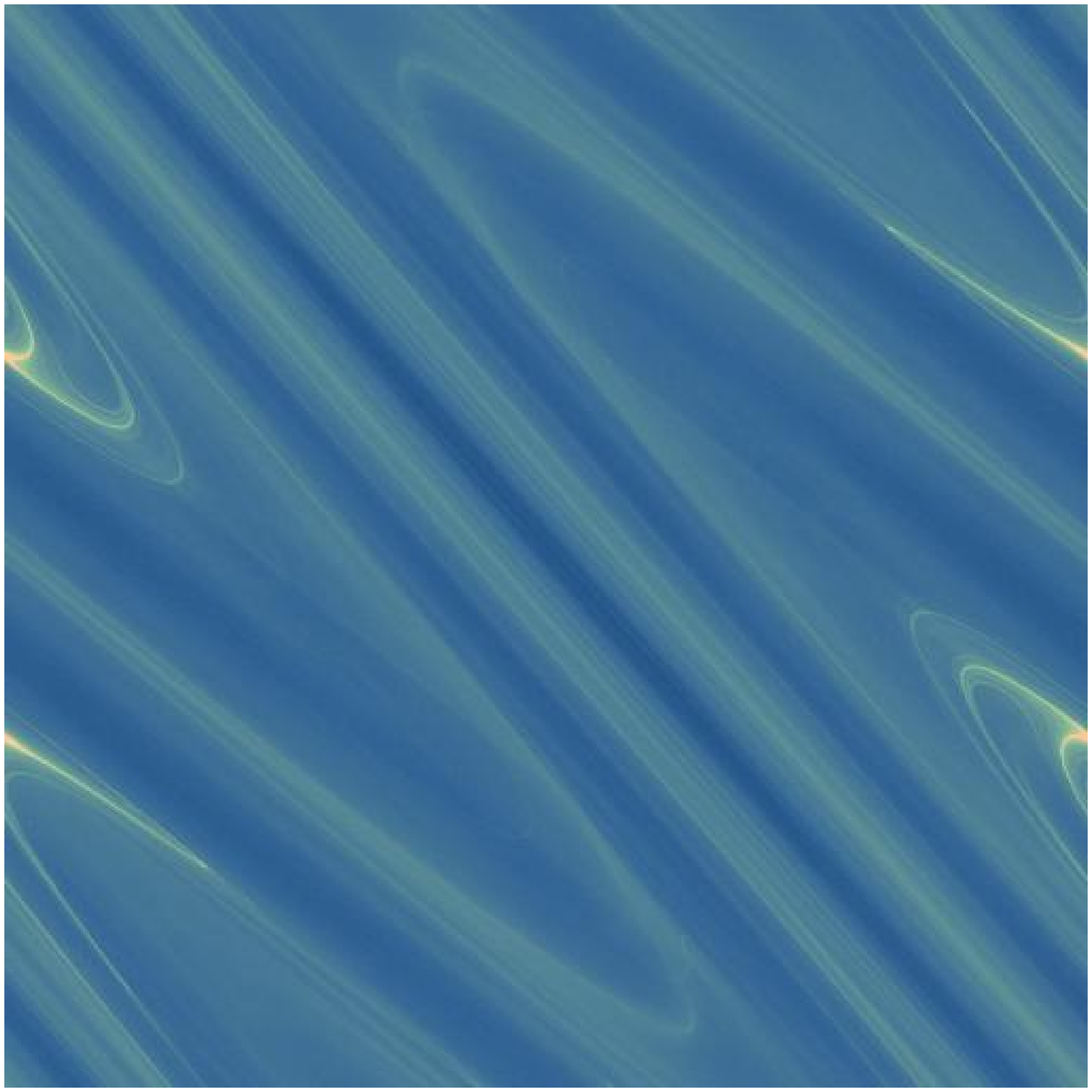} \\
\includegraphics[width=0.49\columnwidth]{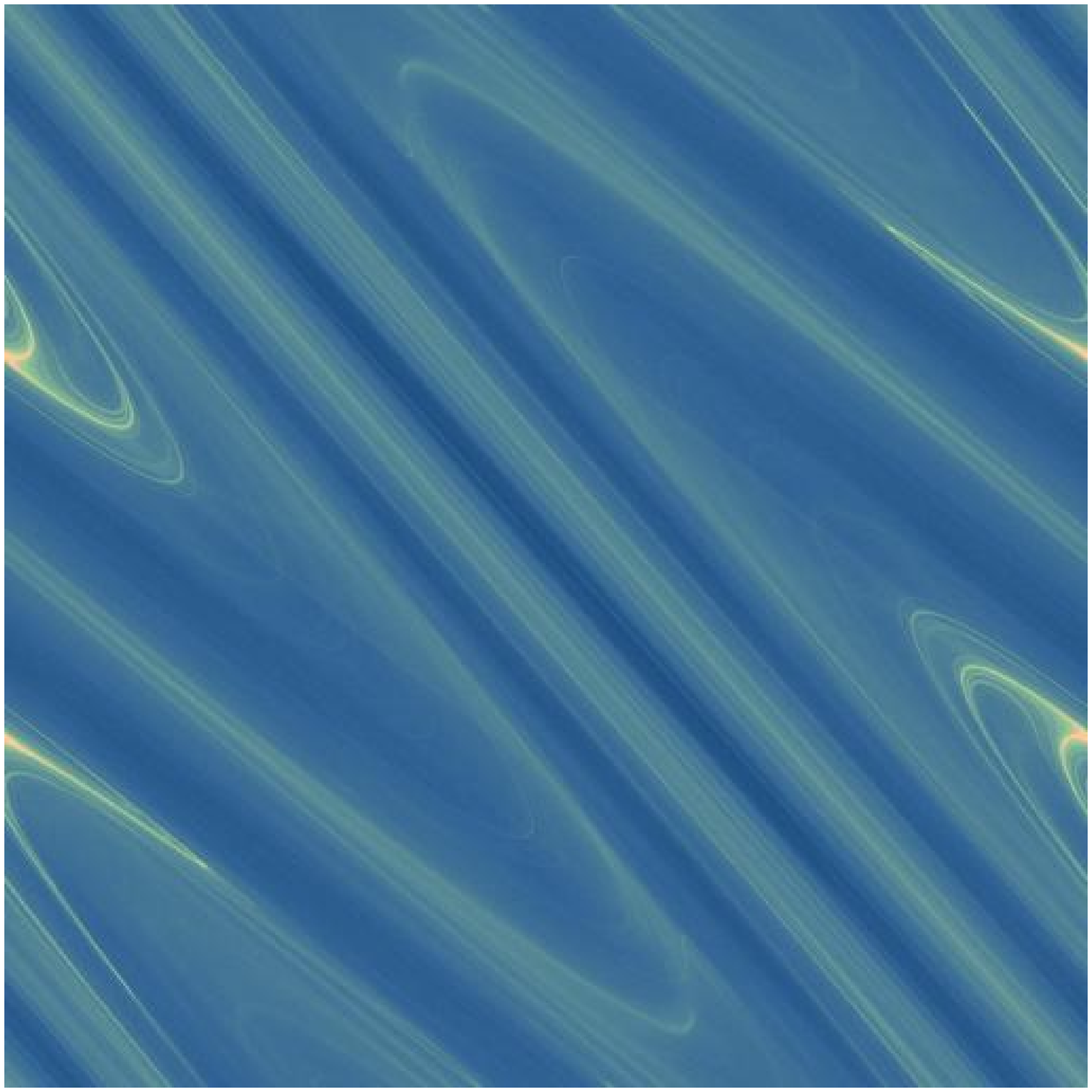}
\includegraphics[width=0.49\columnwidth]{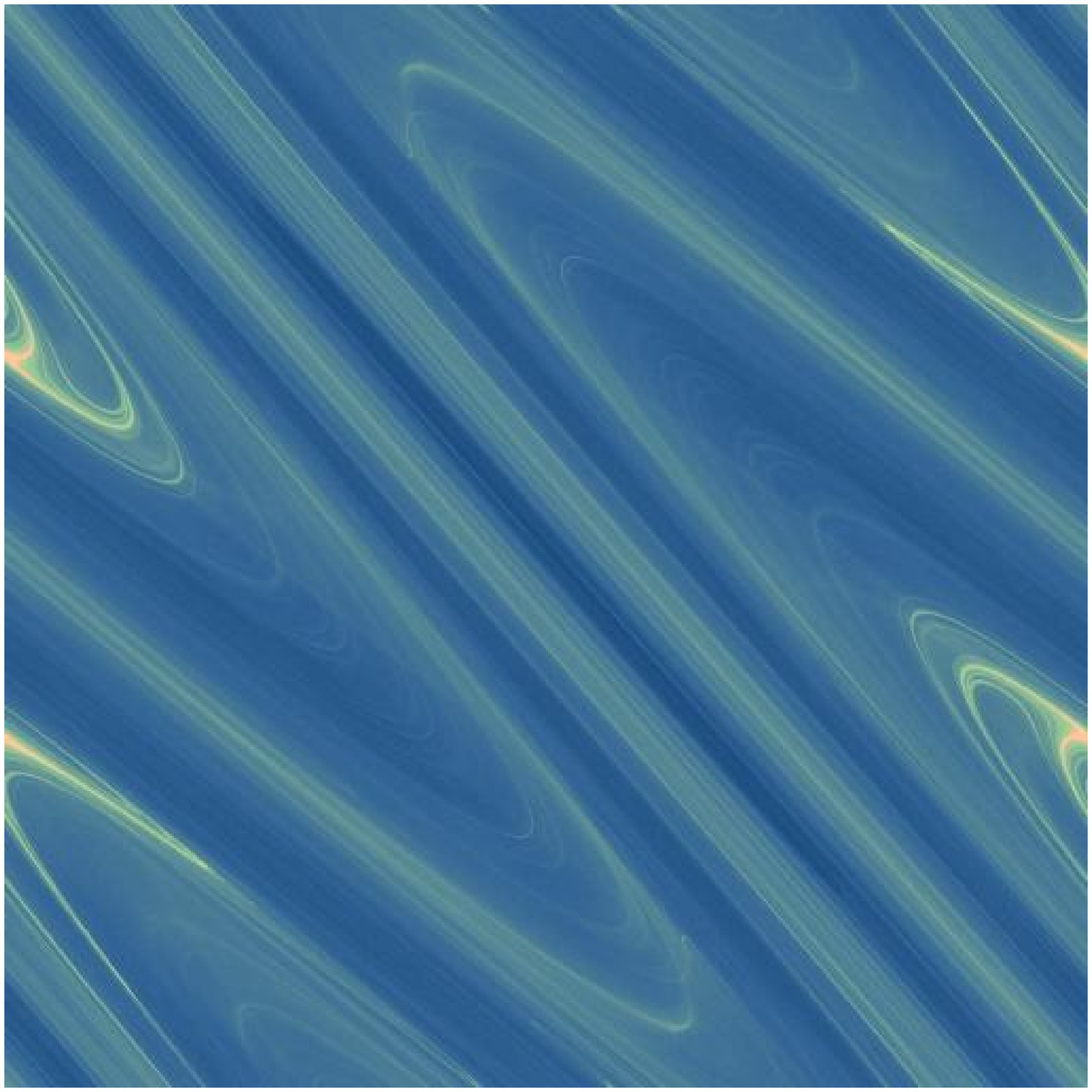}
\end{center}
\caption{\label{out_7}
Histograms with intensity encoding as the square root of invariant probability.
Parameters are standard map parameter 
$k=7.0$, $\varepsilon=10^{-9}$, and, from top left to bottom right,
stability parameter
$\gamma=0.5$, $0.6$, $0.7$, $0.8$, $0.9$, and $1.0$.
}
\end{figure}

Figure~\ref{out_7} displays a sequence of these histograms in a scaled color
code for the same nonlinearity parameter $k=7$ as in Fig.~\ref{k=7},
corresponding to the extremely chaotic regime of the standard map considered in
the previous Section. The sequence of images corresponds to increasing the
stability parameter $\exp(-\gamma)$. The images make evident the fine
filamentary structure developed by the asymptotically invariant distributions
due to the combined effect of noise and the ability of particles to separate
from the basic flow. Remarkably, however, these structures appear even in the
case in which the $\gamma$ values are larger than those required to produce a
spontaneous detachment of the particle trajectories.

Notice that the filamentation here arises from the existence of avenues in the
phase space that lead to the small KAM islands on which the particles prefer to
stay. A more detailed analysis shows that on these avenues the average
value of the squared separation between particles and fluid trajectories is
relatively small. Roughly parallel to these avenues, on the other hand, there
are strips of the phase space that the particles avoid. There, the separation
between particle and fluid trajectories is on average much larger.
Filamentation is thus due to the tendency of the particles to avoid neighboring
regions.

\begin{figure}[tbhp]
\begin{center}
\includegraphics[width=0.95\columnwidth]{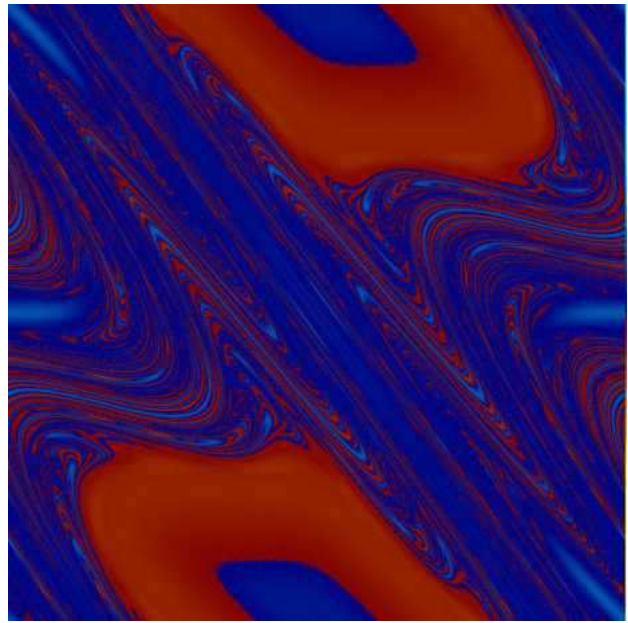}
\end{center}
\caption{\label{marcelo}
Histogram with standard map parameter 
$k=2.0$, $\varepsilon=10^{-7}$, and
stability parameter $\gamma=0.8$.
}
\end{figure}

The same mechanism may also lead to patchiness, not necessarily filamentary.
Looking at flows with weaker chaos, for example, for which relatively large KAM
islands coexist with comparably sized regions of chaos, a situation such as the
one shown in Fig.~\ref{marcelo} is typical. There there are relatively small
avoided regions that separate large patches of larger concentration of
particles around the nonchaotic islands. This picture is also testimonial to a
property that distinguishes the present mechanism from those of Refs.
\cite{jung,toroczkai}. While there the particles group around the unstable
manifolds of the homoclinic intersections of the basic flows, in our case
instead the impurity dynamics tend to avoid the invariant manifolds. This is
because following these manifolds would mean hitting eventually regions in
which the Jacobian eigenvalues are closer to one, which locally amplifies the
effect of fluctuations on the dynamics of the particles.

\section{Discussion and Conclusions}\label{concs}

It has previously been demonstrated that small neutrally buoyant particles
immersed in a fluid flow, and therefore subject to drag forces, may follow
trajectories that spontaneously separate from those of the fluid parcels in
some regions of the flow. Specifically, this occurs when the shear is very
strong compared to the Stokes number associated with the particles. Given that
in general, the smaller the particles are, the stronger is the shear necessary
for this phenomenon to manifest itself, the actual condition for separation may
not be fulfilled by some fluid flows of physical interest. However, we have
shown here that the addition of noise to the forces acting on the particles may
extend the consequences of this phenomenon beyond the range of Stokes numbers
for which separation is possible in an interesting way, namely, the generation
of inhomogeneities in the asymptotic distributions of these particles even in
cases where the flow is a highly efficient mixer.

There is a large variety of examples in which noise ought naturally to be 
added to the dynamical equations of the particles. Thermal fluctuations, for
example should be considered in a range of small-scale laboratory experiments.
The effect of small-scale turbulence forcing drifters in oceanographic
applications might also be another relevant example. In the case of plankton
dynamics, in which the arisal of inhomogeneous distributions is an issue, the
autonomous swimming abilities of individuals might be viewed as an internal
source of noise. But could the phenomenon described here be the basis of the
plankton distribution patchiness? If each individual plankton is naively
considered as an impurity particle, the answer is obviously not: their size is
far too small for these effects to be appreciable. But if there are grounds to 
consider plankton in large-scale colonies moving more or less rigidly
in the ocean, the consequences of the phenomenon described here need to be
taken seriously.

We conclude with an epistemological note: the dynamics of neutral particles in
flows has suggested to us a generalization to maps that helps to solve some
problems in the general domain of Hamiltonian dynamics. This generalization in
turn pays us back by suggesting a way in which inhomogeneous distributions may
arise in fluid-dynamical problems. We believe that this is a remarkable
instance OF mutual convenience in the interdisciplinary marriage between the 
two fields.

\section*{Acknowledgements}

We should like to thank Leo Kadanoff and Marcelo Viana for useful discussions. 
JHEC acknowledges the financial support of the Spanish CSIC, Plan Nacional del
Espacio contract ESP98-1347. MOM acknowledges the support of the Meyer
Foundation. OP acknowledges the Spanish Ministerio de Ciencia y Tecnologia,
Proyecto CONOCE, contract BFM2000-1108. 

\bibliographystyle{prsty}
\bibliography{database}

\begin{thebibliography}{10}

\bibitem{abraham2}
E.~R. Abraham, Nature {\bf 391},  577  (1998).

\bibitem{levin}
S.~A. Levin and L.~A. Segal, Nature {\bf 259},  659  (1976).

\bibitem{steele}
{\em Spatial Patterns in Plankton Communities}, edited by H.~J. Steele (Plenum,
  1978).

\bibitem{jung}
C. Jung, T. T\'el, and E. Ziemniak, Chaos {\bf 3},  555  (1993).

\bibitem{toroczkai}
Z. Toroczkai {\it et~al.}, Phys. Rev. Lett. {\bf 80},  500  (1998).

\bibitem{neufeld}
Z. Neufeld, C. L\'opez, and P.~H. Haynes, Phys. Rev. Lett. {\bf 82},  2606
  (1999).

\bibitem{partproc}
A. Babiano, J.~H.~E. Cartwright, O. Piro, and A. Provenzale,  in {\em Coherent
  Structures in Complex Systems}, Vol.~567 of {\em Lecture Notes in Physics},
  edited by D. Reguera, L. Bonilla, and M. Rubi (Springer, 2001), pp.\
  114--126.

\bibitem{neutpartprl}
A. Babiano, J.~H.~E. Cartwright, O. Piro, and A. Provenzale, Phys. Rev. Lett.
  {\bf 84},  5764  (2000).

\bibitem{feingold88.2}
M. Feingold, L.~P. Kadanoff, and O. Piro, J. Stat. Phys. {\bf 50},  529
  (1988).

\bibitem{pirofein}
O. Piro and M. Feingold, Phys. Rev. Lett. {\bf 61},  1799  (1988).

\bibitem{3dpaper}
J.~H.~E. Cartwright, M. Feingold, and O. Piro, Physica D {\bf 76},  22  (1994).

\bibitem{spherespaper}
J.~H.~E. Cartwright, M. Feingold, and O. Piro, J. Fluid Mech. {\bf 316},  259
  (1996).

\bibitem{spheresletter}
J.~H.~E. Cartwright, M. Feingold, and O. Piro, Phys. Rev. Lett. {\bf 75},  3669
   (1995).

\bibitem{abel}
M. Abel, A. Celani, D. Vergni, and A. Vulpiani, Phys. Rev. E {\bf 64},  046307
  (2001).

\bibitem{lopez}
C. L\'opez {\it et~al.}, Chaos {\bf 11},  397  (2001).

\bibitem{bailout1}
J.~H.~E. Cartwright, M.~O. Magnasco, and O. Piro, submitted  (2001).

\bibitem{maxey}
M.~R. Maxey and J.~J. Riley, Phys. Fluids {\bf 26},  883  (1983).

\bibitem{michaelides}
E.~E. Michaelides, J. Fluids Eng. {\bf 119},  233  (1997).

\bibitem{taylor}
G.~I. Taylor, Proc. Roy. Soc. Lond. A {\bf 120},  260  (1928).

\bibitem{auton}
T.~R. Auton, J.~C.~R. Hunt, and M. Prud'homme, J. Fluid Mech. {\bf 197},  241
  (1988).

\bibitem{boussinesq}
J. Boussinesq, C. R. Acad. Sci. Paris {\bf 100},  935  (1885).

\bibitem{basset}
A.~B. Basset, Phil. Trans. Roy. Soc. Lond. {\bf 179},  43  (1888).

\bibitem{faxen}
H. Fax{\'e}n, Ann. Phys. {\bf 4},  89  (1922).

\bibitem{okubo}
A. Okubo, Deep-Sea Res. {\bf 17},  445  (1970).

\bibitem{weiss2}
J.~B. Weiss, Physica D {\bf 48},  273  (1991).

\end{thebibliography}

\end{document}